\documentclass[traditabstract,rnote]{aa}

\usepackage{natbib}
\usepackage{amsmath}
\usepackage{graphicx}
\usepackage{wasysym}
\usepackage{txfonts}
\usepackage{xspace}
\usepackage{url}
\usepackage{rotating}

\newcommand{\fu}{4U\,1909+07\xspace}

\newcommand{\inte}{\textsl{INTEGRAL}\xspace}

\newcommand{\suz}{\textsl{Suzaku}\xspace}
\newcommand{\uhuru}{\textsl{Uhuru}\xspace}
\newcommand{\xte}{\textsl{RXTE}\xspace}

\newcommand{\snr}{S/N\xspace}

\newcommand{\redchi}{\ensuremath{\chi^{2}_\text{red}}}

\newcommand{\feka}{\ensuremath{\mathrm{Fe}~\mathrm{K}\alpha}\xspace}
\newcommand{\nh}{\ensuremath{{N}_\mathrm{H}}\xspace}
\newcommand{\nhone}{\ensuremath{{N}_{\mathrm{H},1}}\xspace}
\newcommand{\nhtwo}{\ensuremath{{N}_{\mathrm{H},2}}\xspace}

\newcommand{\fuerst}{F11\xspace}

\begin{document}

%Title of paper
\title{Staring at \fu with  \suz}

\titlerunning{Staring at \fu with  \suz}
\author{
 \mbox{F. F\"urst\inst{1,2}} \and
\mbox{K. Pottschmidt\inst{3,4}} \and
 \mbox{I. Kreykenbohm\inst{1}} \and
\mbox{S. M\"uller\inst{1}} \and
\mbox{M. K\"uhnel\inst{1}} \and
\mbox{J. Wilms\inst{1}} \and
\mbox{R. E. Rothschild\inst{5}}
}
\authorrunning{F.~F\"urst et al.}

\institute{
 Dr.~Karl Remeis-Sternwarte \& ECAP, Universit\"at Erlangen-N\"urnberg, Sternwartstr.~7, 96049~Bamberg, Germany
\and Space Radiation Lab, California Institute of Technology, MC 290-17 Cahill, 1200 E. California Blvd., Pasadena, CA 91125, USA
\and CRESST and NASA Goddard Space Flight Center, Astrophysics Science Division, Code 661, Greenbelt, MD 20771, USA
\and Center for Space Science and Technology, University of Maryland Baltimore County, 1000 Hilltop Circle, Baltimore, MD 21250, USA
\and Center for Astrophysics \& Space Sciences, University of California, San Diego, 9500 Gilman Drive, La Jolla, CA 92093, USA
}
\date{Received: --- / Accepted: ---}

\abstract{
We present an analysis of the neutron star High Mass X-ray Binary (HMXB) \fu mainly based  on  \suz data. We extend the pulse period evolution, which behaves in a random-walk like manner, indicative of direct wind accretion. Studying the spectral properties of \fu between 0.5 to 90\,keV we find that a power-law with an exponential cutoff can describe the data well, when additionally allowing for a blackbody or a partially covering absorber at low energies. 

We find no evidence for a cyclotron resonant scattering feature (CRSF), a feature seen in many other neutron star HMXBs sources. By performing pulse phase resolved spectroscopy we investigate the origin of the strong energy dependence of the pulse profile, which evolves from a broad two-peak profile at low energies to a profile with a single, narrow peak at energies above 20\,keV. Our data show that it is very likely that a higher folding energy in the high energy peak is responsible for this behavior. This in turn leads to the assumption that we observe the two magnetic poles and their respective accretion columns at different phases, and that these accretions column have slightly different physical conditions.
}
\keywords{stars: neutron (4U 1909+07) -- X-rays: binaries -- Accretion }

\maketitle

\section{Introduction}
 
High mass X-ray binaries (HMXBs), in which the compact object is a rotating neutron star, provide a unique way to investigate the flow of matter into a deep gravitational potential. By analyzing the evolution of the pulse period and its correlation to luminosity, the presence or absence of a stable accretion disk can be probed \citep{ghosh79a}. By performing pulse phase resolved spectroscopy, different areas of the X-ray producing region can be analyzed and thereby the interaction between radiation and matter in very strong magnetic fields can be studied. \fu is an ideal source for these investigations, as it is a slowly rotating, persistent HMXB. Discovered by \uhuru \citep{giacconi74a}, its 4.4\,d orbital period was determined in \xte/ASM data by \citet{wen00a}. Pulsations with a pulse period of $\approx$$605$\,s were later found in \xte/PCA data by \citet{levine04a}. These authors also found that the intrinsic absorption \nh is strongly variable over the orbit, peaking around orbital phase 0. These measurements can be explained by a smooth stellar wind and an inclination of the system between 38$^\circ$ and 68$^\circ$ \citep{levine04a}. \citet[hereafter \fuerst]{fuerst10b} investigated the long-term evolution of the pulse period between 2003 and 2009 using \inte/ISGRI lightcurves. We found that the evolution is consistent with a random walk, a typical behavior for a wind accreting source without a stable accretion disk, similar to \mbox{Vela~X-1} \citep[see, e.g.,][]{dekool93a}.

The spectrum of \fu can be described using typical phenomenological models often applied to HMXB neutron star sources, such as a power-law with an exponential cutoff \citep[\fuerst]{levine04a}. By analyzing  phase-resolved \xte data we found that a blackbody component with a temperature around 1.7\,keV is also needed to describe the soft excess in the spectrum. Many neutron star sources show a cyclotron resonant scattering feature \citep[CRSF, see, e.g.,][]{schoenherr07a} at energies between 10 and 50\,keV. Neither in \xte/HEXTE nor in \inte/ISGRI data is evidence for such a feature seen; however, the signal-to-noise ratio (\snr) in these data is rather low. To investigate the behavior of the soft excess in more detail and to perform a rigorous search for a CRSF, we obtained a 20\,ks \suz observation, the analysis of which is presented in this article.
 In Sect.~\ref{sec:obs} we give a short overview of the data used and the reduction pipeline. In Sect.~\ref{sec:timing} we discuss the evolution of the pulse period, while in Sect.~\ref{sec:spec} we analyze the X-ray spectrum. We perform phase averaged (Sect.~\ref{susec:phasavg_spec}) as well as phase resolved spectroscopy (Sect.~\ref{susec:phasres_spec}). In Sect.~\ref{sec:summary} we summarize and discuss our results.

\section{Observations \& data reduction}
\label{sec:obs}
\begin{figure}
 \includegraphics[width=0.95\columnwidth]{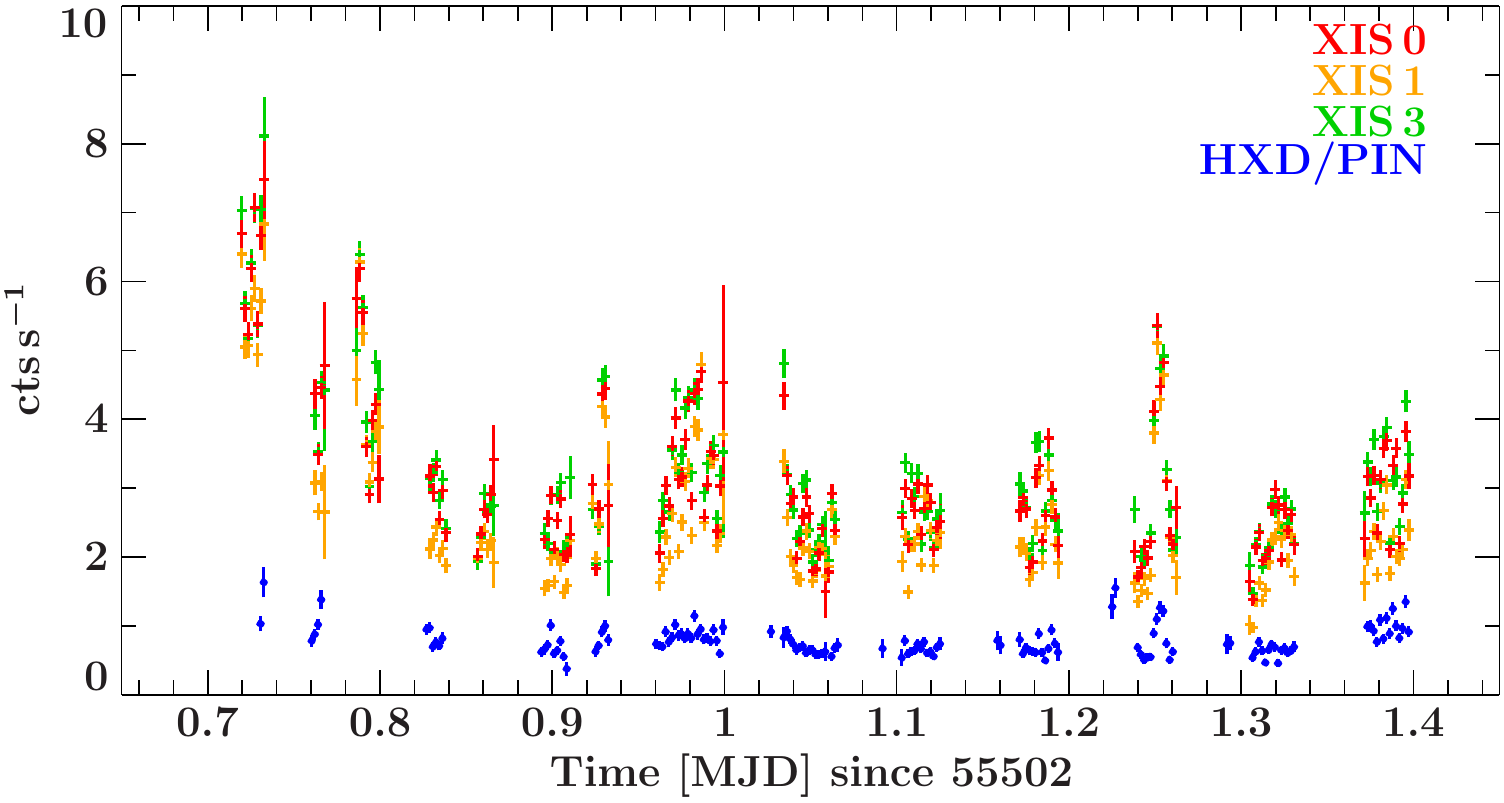}
 \caption{Lightcurves taken with XIS\,0 (red), XIS\,1 (orange), and XIS\,3 (green) in the 0.5--10\,keV energy band and PIN (blue) between 10 to 20\,keV, rebinned to 160\,s.}
 \label{fig:lc}
\end{figure}

This paper is mainly focused on a 20\,ks \suz observation performed on November 2, 2010. Data were taken in the orbital phase range 0.472--0.627, i.e., outside the phases of increased absorption. We use data from the X-ray Imaging Spectrometers \citep[XISs,][]{koyama07a} 0, 1, and 3 in the energy range 0.5--10\,keV and from the High X-ray Detector's \citep[HXD, ][]{takahashi07a} PIN instrument from 10\,keV to 90\,keV. We did not use data from the HXD/GSO due to source confusion with the nearby, bright black-hole binary GRS~1915+105. As the hard X-ray spectrum of \fu falls off exponentially, a significant detection in GSO would be rather unlikely. The data were reduced using the standard pipeline from HEASOFT v6.11 and the HXD calibration from September 2011, including deadtime correction for PIN. Data analysis was performed using the Interactive Spectral Interpretation System \citep[ISIS, ][]{houck00a}, version 1.6.2. We extracted lightcurves from XIS with 16\,s time resolution in the energy bands 0.5--5\,keV, 5--10\,keV, and 0.5--10\,keV, as well as from PIN with 40\,s resolution in the energy bands 10--20\,keV, 20--40\,keV, and 40--90\,keV. Figure~\ref{fig:lc} shows the lightcurves from all instruments, rebinned to 160\,s for better visibility. Some flaring is evident in the lightcurve, however, no significant changes in hardness could be found when comparing the different energy bands. 
For comparison we also extracted data from the ISGRI detector \citep{lebrun03a} aboard \inte \citep{winkler03a}, between MJD 55107.13 to 55148.82 and MJD 55503.74 to 55518.02 with exposure times of 235\,ks and 396\,ks, respectively. Using the standard OSA~9.0 we extracted lightcurves between 20 to 40\,keV with 20\,s time resolution. All lightcurves, \suz and \inte, were transferred to the barycenter of the solar system and corrected for the orbital motion of \fu, using the ephemeris by \citet{levine04a}. 

\section{Timing}
\label{sec:timing}
\begin{figure}
 \includegraphics[width=0.95\columnwidth]{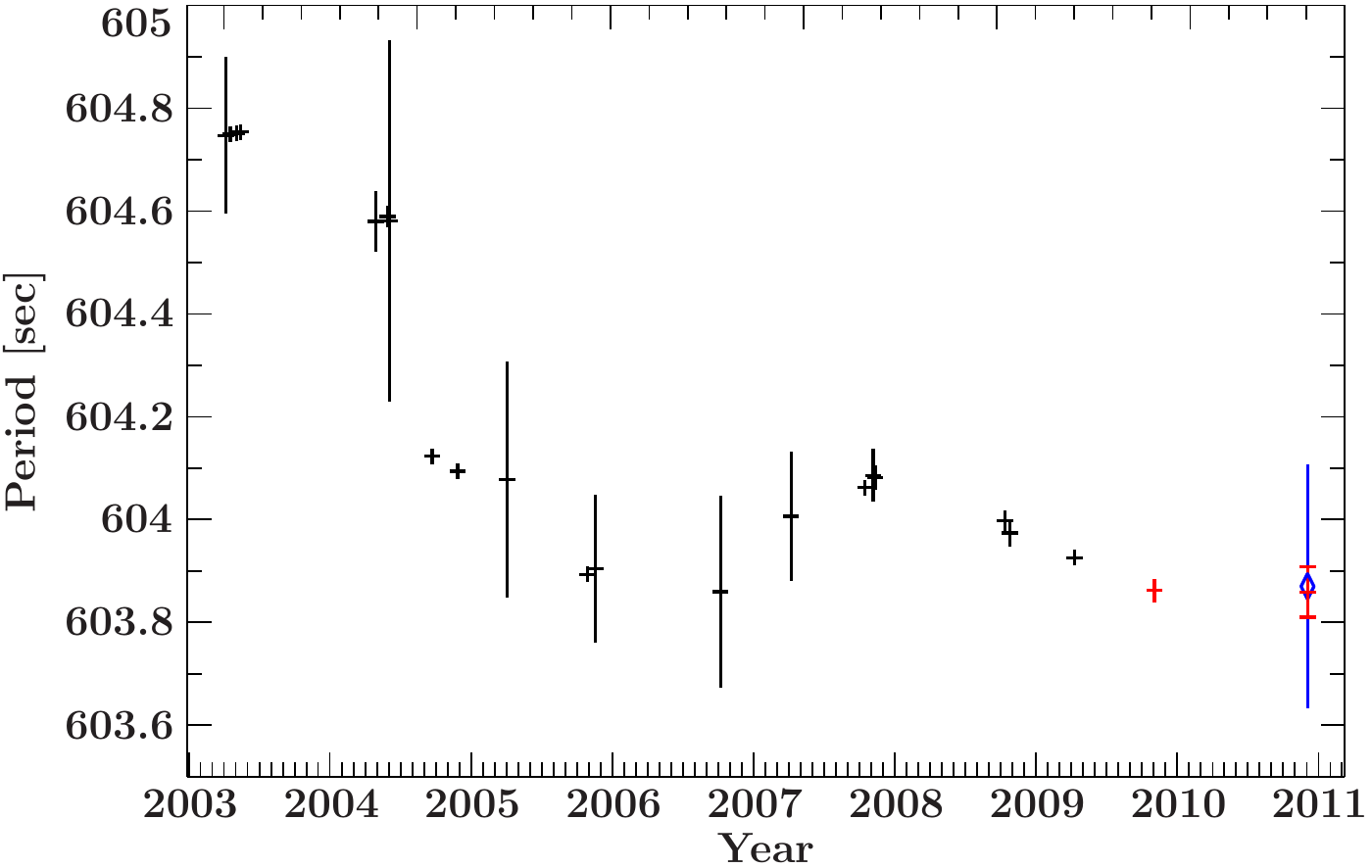}
 \caption{Pulse period evolution between 2003--2011. Black data were already published by \fuerst. More recent \inte/ISGRI data are shown in red, \suz/PIN in blue. \suz/XIS is not shown to ensure readability.}
 \label{fig:pulsevo}
\end{figure}

The  pulse period evolution of \fu was published by \fuerst up to  spring 2009. Figure~\ref{fig:pulsevo} extends this time series up to fall 2010, using all publicly available \inte data and the \suz observation. \inte also observed \fu in spring 2010, but the rough sampling does not allow for a reliable pulse period measurement. To obtain the pulse periods we performed epoch folding \citep{leahy83b} using XIS lightcurves between 0.5--10\,keV with 16\,s time resolution, PIN lightcurves between 10--90\,keV with 40\,s time resolution, and \inte/ISGRI lightcurves between 20--40\,keV with 20\,s time resolution.
Uncertainties on the pulse periods of \inte were obtained by fitting a Gaussian distribution to the $\chi^2$ distribution of the individual epoch folding results, as described by \fuerst. The epoch folding distribution of \suz/XIS as well as \suz/PIN does not show a well enough defined peak for fitting with a Gaussian function. We therefore estimated the uncertainties on the period measurement using a Monte Carlo approach as described by \citet{davies90a}, where we simulated 1000 lightcurves with the same binning as the original measurement, scattered around the measured pulse profile. The width of the distribution of resulting pulse periods in these lightcurves is a good indicator of the uncertainty of the pulse period. 
This approach leads to $P_\text{XIS}=604.10\pm0.20$\,s and $P_\text{PIN} = 603.87\pm0.24$\,s (all uncertainties are given at the 90\% level). Due to the longer exposure time, the almost simultaneous \inte/ISGRI data provide a much better constrained pulse period measurement, with $P_\text{ISGRI} = 603.86\pm0.05$\,s. As this period is consistent with the \suz measurements, we will use it throughout this paper and for the analysis of the \suz data.

\begin{figure}
 \includegraphics[width=0.95\columnwidth]{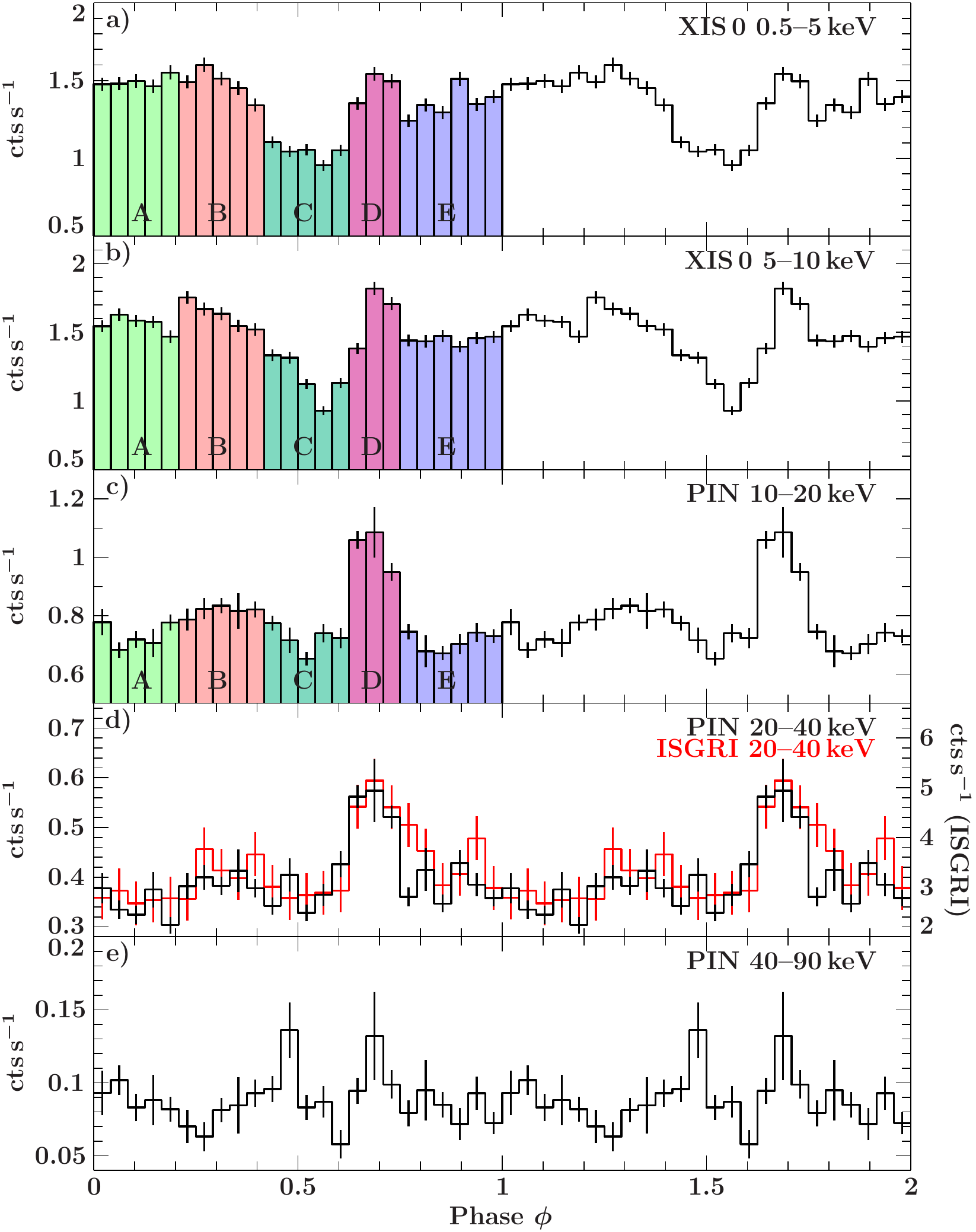}
\caption{Energy resolved pulse profiles as measured with \suz/XIS (\textit{a)} and \textit{b)}) and PIN (\textit{c)} to \textit{e)}). Marked in color and labeled A--E are the phase bins used for phase-resolved spectroscopy (see text for details). In panel \textit{c)}, \inte/ISGRI data is superimposed in red, scaled according the right-hand $y$-axis.}
 \label{fig:ppergres}
\end{figure}

The pulse profiles of \suz/PIN and \inte/ISGRI in the 20--40\,keV range agree, showing a short single peak followed by a long dim phase (Fig.~\ref{fig:ppergres}d). This profile is in agreement with \xte/HEXTE \citep[see, e.g., ][\fuerst]{levine04a}. As described by \fuerst a strong change in the pulse profile occurs between 8 and 20\,keV, where it changes from a two-peaked profile to a single peaked profile. This energy dependent behavior is also seen in other neutron star X-ray binaries, like 4U~0115+634 \citep{mueller10a} and 1A~1118$-$61 \citep{suchy11b}. The  energy range of the transition is only marginally covered by the instruments aboard \suz, as it is just in the overlapping region between XIS and PIN. The XIS pulse profile in different energy bands is shown in Fig.~\ref{fig:ppergres}a and Fig.~\ref{fig:ppergres}b and is very similar to \xte/PCA profiles as presented by \citet{levine04a} and \fuerst. 

\section{Spectroscopy}
\label{sec:spec}
\subsection{Phase-averaged spectroscopy}
\label{susec:phasavg_spec}

\begin{figure}
 \includegraphics[width=0.95\columnwidth]{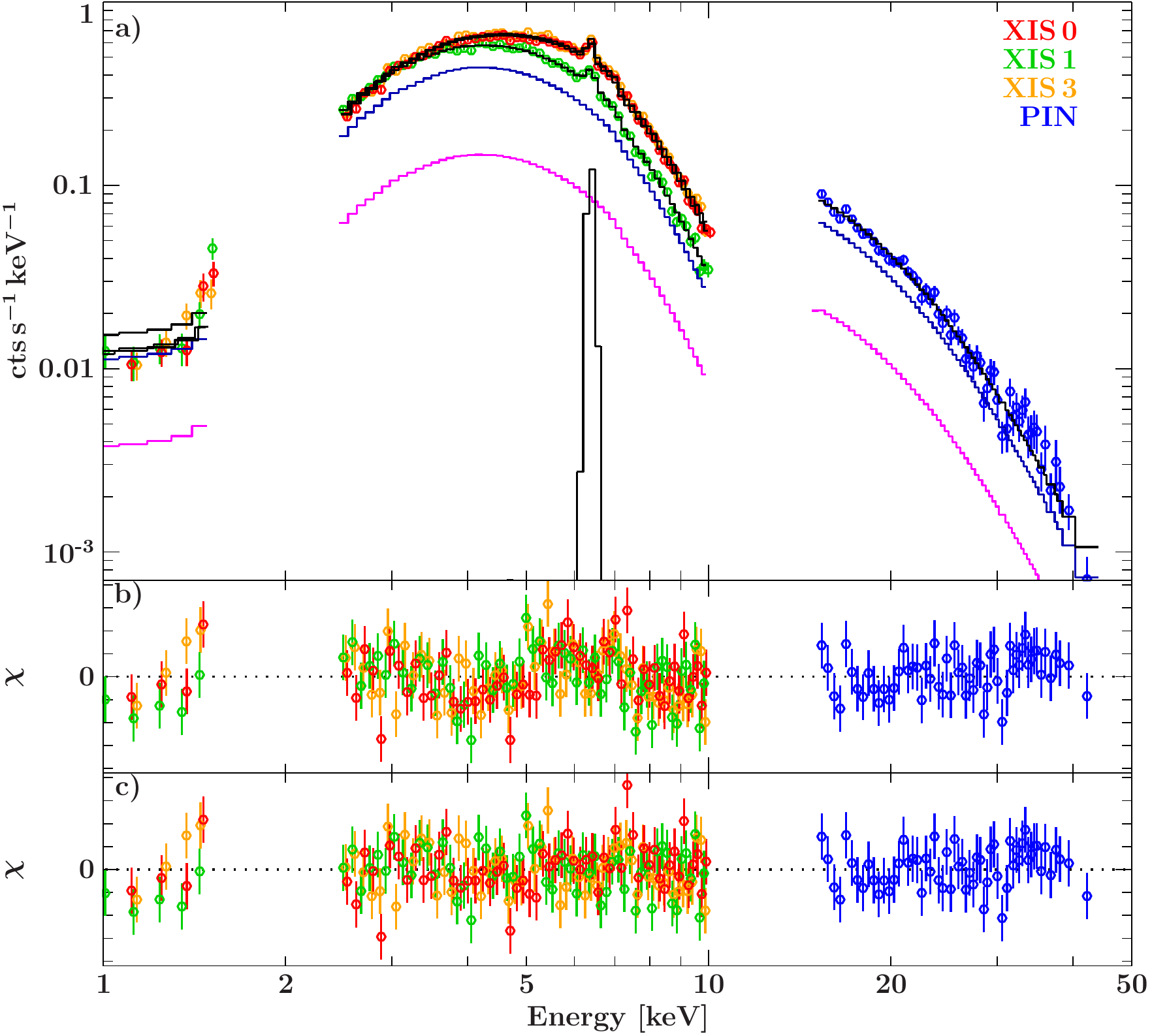}
 \caption{\textit{a)} XIS0 (red), XIS1 (green), XIS3 (orange), and PIN (blue) spectrum together with the best fit \texttt{cutoffpl}-model including partial covering. In magenta the strongly absorbed part is shown, in dark blue the less absorbed one. The \feka line is shown in brown. \textit{b)} residuals to the \texttt{cutoffpl} model without partial coverer. \textit{c)} residuals to the best-fit model.}
 \label{fig:avgspec}
\end{figure}

As discussed by \fuerst, the phase averaged spectrum of \fu can be very well described with typical phenomenological models often used to describe the spectra of neutron stars in HMXBs plus a Gaussian-shaped fluorescent \feka line. The XSPEC models \texttt{cutoffpl}, \texttt{highecut} \citep{white83a}, and \texttt{comptt} \citep{hua95a} all gave similar statistically acceptable fits. Therefore, we applied these models to the \suz data and the results are shown in  Tables~\ref{tab:phasavg_ecu}, \ref{tab:phasavg_hie}, and \ref{tab:phasavg_ctt}, respectively. The data are shown together with the best-fit \texttt{cutoffpl} model in Fig.~\ref{fig:avgspec}.

The flux of the \suz data (XIS normalization\footnote{PIN cross-calibration factor was within 2\%}) is comparable to the \xte  data (PCA normalization) presented by \fuerst, with $\mathcal{F}_\text{\suz}^{\text{0.5--100\,keV}}=0.324^{+0.019}_{-0.017}$\,keV\,s$^{-1}$\,cm$^{-2}$ and $\mathcal{F}_\text{\xte}^{\text{0.5--100\,keV}}=0.330\pm0.007$\,keV\,s$^{-1}$\,cm$^{-2}$, respectively.
The \feka line is narrow and its width could not be resolved in the XIS data. We therefore fixed it to  $10^{-6}$\,keV in all fits.
It was shown by \fuerst that an additional blackbody component is necessary to describe a soft excess. Adding this component leads also to a clear improvement in the fit to the \suz data, as can be seen in the second column of Tables~\ref{tab:phasavg_ecu}, \ref{tab:phasavg_hie}, and \ref{tab:phasavg_ctt}. However, some continuum parameters become only very weakly constrained, like $E_\text{fold}$ in the \texttt{cutoffpl} and \texttt{highecut} model or the seed photon temperature $T_0$ in the \texttt{comptt} model. Additionally the 20-40\,keV flux drops to unrealistically low values\footnote{compensated by a very high PIN cross-calibration factor of $\approx$$1.3$} and the best fit black body temperature is very different between the  \texttt{cutoffpl} and \texttt{highecut} model, rendering a physical interpretation impossible. 
To describe the soft energy part of the spectrum without a black body component, we included a partial coverer in the model with two different absorption columns \nhone and \nhtwo, with their relative influence determined by a covering fraction CF:
\begin{equation}
 (\mathrm{CF}\times\nhone + (1-\mathrm{CF}) \times  \nhtwo ) \times ( \texttt{cont} + \feka )
\end{equation}
 This model leads to equally large improvements in terms of $\chi^2$, as seen in the third column of Tables~\ref{tab:phasavg_ecu}, \ref{tab:phasavg_hie}, and \ref{tab:phasavg_ctt}. Although the secondary absorption column \nhtwo is also only weakly constrained, the continuum parameters remain very well constrained and at physically reasonable values. 
 
\begin{table}
\caption{Phase averaged spectral parameters for the \texttt{cutoffpl} continuum model. In the second column the best-fit parameters  with an additional blackbody component (bbody) are shown, in the third the parameters when including a partial coverer (PC). }
\label{tab:phasavg_ecu}
 \begin{tabular}{lccc}
\hline\hline
Model               & cutoffpl  & cutoffpl   & cutoffpl  \\
parameter           &     & +bbody  & +PC \\\hline
$\mathcal{F}_\text{3--10\,keV} $\tablefootmark{a} & $9.33\pm0.07$ & $9.27\pm0.07$ & $9.29\pm0.07$ \\
$\mathcal{F}_\text{20--40\,keV}$\tablefootmark{a} & $8.6\pm0.7$ & $6.9^{+0.9}_{-0.8}$ & $8.7\pm0.7$ \\
$N_\text{H}$-1\tablefootmark{b} & $10.71\pm0.30$ & $9.7^{+0.6}_{-0.4}$ & $51^{+16}_{-15}$ \\
$N_\text{H}$-2\tablefootmark{b} & -- & -- & $10.8^{+0.5}_{-0.7}$ \\
CF\tablefootmark{c} & -- & -- & $0.25\pm0.06$ \\
$\Gamma$ & $0.93\pm0.06$ & $1.09^{+0.17}_{-0.10}$ & $1.20\pm0.09$ \\
$E_\text{fold}$\,[keV] & $15.4^{+1.7}_{-1.5}$ & $22^{+8}_{-4}$ & $19.3^{+3.0}_{-2.4}$ \\
$E_\text{Fe}$\,[keV] & $6.388^{+0.016}_{-0.005}$ & $6.388^{+0.016}_{-0.005}$ & $6.391^{+0.012}_{-0.009}$ \\
$A_\text{Fe}$\tablefootmark{d} & $1.98\pm0.18$ & $1.84\pm0.18$ & $1.94\pm0.20$ \\
$kT_\text{bb}$\,[keV] & -- & $2.03^{+0.19}_{-0.22}$ & -- \\
$A_\text{bb}$ & -- & $\left(0.98^{+0.28}_{-0.29}\right)\times10^{-3}$ & -- \\
$\chi^2$/dof  &  544.78/414 & 506.74/412 & 497.76/412  \\
\end{tabular}
\tablefoot{\tablefoottext{a}{in $10^{-2}$\,keV\,s$^{-1}$\,cm$^{-2}$} \tablefoottext{b}{in units of $10^{22}$\,H-atoms\,cm$^{-2}$}
\tablefoottext{c}{Covering Fraction} \tablefoottext{d}{in $10^{-4}$\,ph\,s$^{-1}$\,cm$^{-2}$}.}
\end{table}

\begin{table}
\caption{Same as Tab.~\ref{tab:phasavg_ecu}, but for the \texttt{highecut} model.}
\label{tab:phasavg_hie}
 \begin{tabular}{lccc}
\hline\hline
Model               & highecut  & highecut   & highecut  \\
parameter           &     & +bbody  & +PC \\\hline
$\mathcal{F}_\text{3--10\,keV} $\tablefootmark{a} & $9.28\pm0.07$ & $9.27\pm0.07$ & $9.28\pm0.07$ \\
$\mathcal{F}_\text{20--40\,keV}$\tablefootmark{a} & $8.1\pm0.7$ & $8.0^{+0.9}_{-0.7}$ & $8.6^{+0.8}_{-1.3}$ \\
$N_\text{H}$-1\tablefootmark{b} & $10.3\pm0.4$ & $14.5^{+2.3}_{-3.9}$ & $90^{+100.0}_{-70}$ \\
$N_\text{H}$-2\tablefootmark{b} & -- & -- & $10.4^{+1.2}_{-0.4}$ \\
CF\tablefootmark{c} & -- & -- & $0.15^{+0.08}_{-0.10}$ \\
$\Gamma$ & $1.12^{+0.06}_{-0.07}$ & $1.38^{+0.10}_{-0.16}$ & $1.19^{+0.34}_{-0.09}$ \\
$E_\text{fold}$\,[keV] & $18.0^{+1.9}_{-1.6}$ & $24^{+5}_{-7}$ & $19.05^{+0.11}_{-2.45}$ \\
$E_\text{cut}$\,[keV] & $6.0\pm0.4$ & $7.0^{+0.6}_{-1.7}$ & $5.5^{+2.3}_{-0.4}$ \\
$E_\text{Fe}$\,[keV] & $6.390^{+0.014}_{-0.007}$ & $6.3839^{+0.0191}_{-0.0009}$ & $6.390^{+0.014}_{-0.007}$ \\
$A_\text{Fe}$\tablefootmark{d} & $1.82\pm0.19$ & $1.83^{+0.20}_{-0.24}$ & $1.96^{+0.27}_{-0.31}$ \\
$kT_\text{bb}$\,[keV] & -- & $0.34^{+0.12}_{-0.10}$ & -- \\
$A_\text{bb}$ & -- & $\left(2.3^{+2.6}_{-2.2}\right)\times10^{-3}$ & -- \\
$\chi^2$/dof  &  499.20/413 & 494.65/411 & 492.03/411  \\
\end{tabular}
\tablefoot{\tablefoottext{a}{in $10^{-2}$\,keV\,s$^{-1}$\,cm$^{-2}$} \tablefoottext{b}{in units of $10^{22}$\,H-atoms\,cm$^{-2}$}
\tablefoottext{c}{Covering Fraction} \tablefoottext{d}{in $10^{-4}$\,ph\,s$^{-1}$\,cm$^{-2}$}.}
\end{table}

\begin{table}
\caption{Same as Tab.~\ref{tab:phasavg_ecu}, but for the \texttt{compTT} model.}
\label{tab:phasavg_ctt}
 \begin{tabular}{lccc}
\hline\hline
Model               & compTT  & compTT   & compTT  \\
parameter           &     & +bbody  & +PC \\\hline
$\mathcal{F}_\text{3--10\,keV} $\tablefootmark{a} & $9.27\pm0.07$ & $9.27\pm0.07$ & $9.27\pm0.07$ \\
$\mathcal{F}_\text{20--40\,keV}$\tablefootmark{a} & $8.6\pm0.8$ & $6.7^{+0.9}_{-0.8}$ & $9.4^{+0.9}_{-0.8}$ \\
$N_\text{H}$-1\tablefootmark{b} & $6.3\pm0.4$ & $10.3^{+0.7}_{-1.0}$ & $86^{+14}_{-22}$ \\
$N_\text{H}$-2\tablefootmark{b} & -- & -- & $7.1^{+0.7}_{-0.6}$ \\
CF\tablefootmark{c} & -- & -- & $0.28\pm0.06$ \\
$T_0$\,[keV] & $1.18\pm0.06$ & $0.34^{+0.15}_{-0.33}$ & $1.04^{+0.09}_{-0.10}$ \\
$kT$\,[keV] & $7.3^{+0.7}_{-0.5}$ & $8.0^{+1.3}_{-0.9}$ & $7.3^{+0.7}_{-0.5}$ \\
$\tau$ & $4.3\pm0.4$ & $4.4^{+0.8}_{-0.7}$ & $4.1\pm0.4$ \\
$E_\text{Fe}$\,[keV] & $6.388^{+0.015}_{-0.006}$ & $6.388^{+0.015}_{-0.006}$ & $6.387^{+0.016}_{-0.005}$ \\
$A_\text{Fe}$\tablefootmark{d} & $1.87\pm0.17$ & $1.83\pm0.18$ & $1.99^{+0.22}_{-0.20}$ \\
$kT_\text{bb}$\,[keV] & -- & $2.13\pm0.13$ & -- \\
$A_\text{bb}$ & -- & $\left(1.41\pm0.20\right)\times10^{-3}$ & -- \\
$\chi^2$/dof  &  551.26/413 & 501.33/411 & 506.79/411  \\
\end{tabular}
\tablefoot{\tablefoottext{a}{in $10^{-2}$\,keV\,s$^{-1}$\,cm$^{-2}$}  \tablefoottext{b}{in units of $10^{22}$\,H-atoms\,cm$^{-2}$}
\tablefoottext{c}{Covering Fraction} \tablefoottext{d}{in $10^{-4}$\,ph\,s$^{-1}$\,cm$^{-2}$}.}
\end{table}

\subsection{Phase-resolved spectroscopy}
\label{susec:phasres_spec}
\begin{figure}
 \includegraphics[width=0.95\columnwidth]{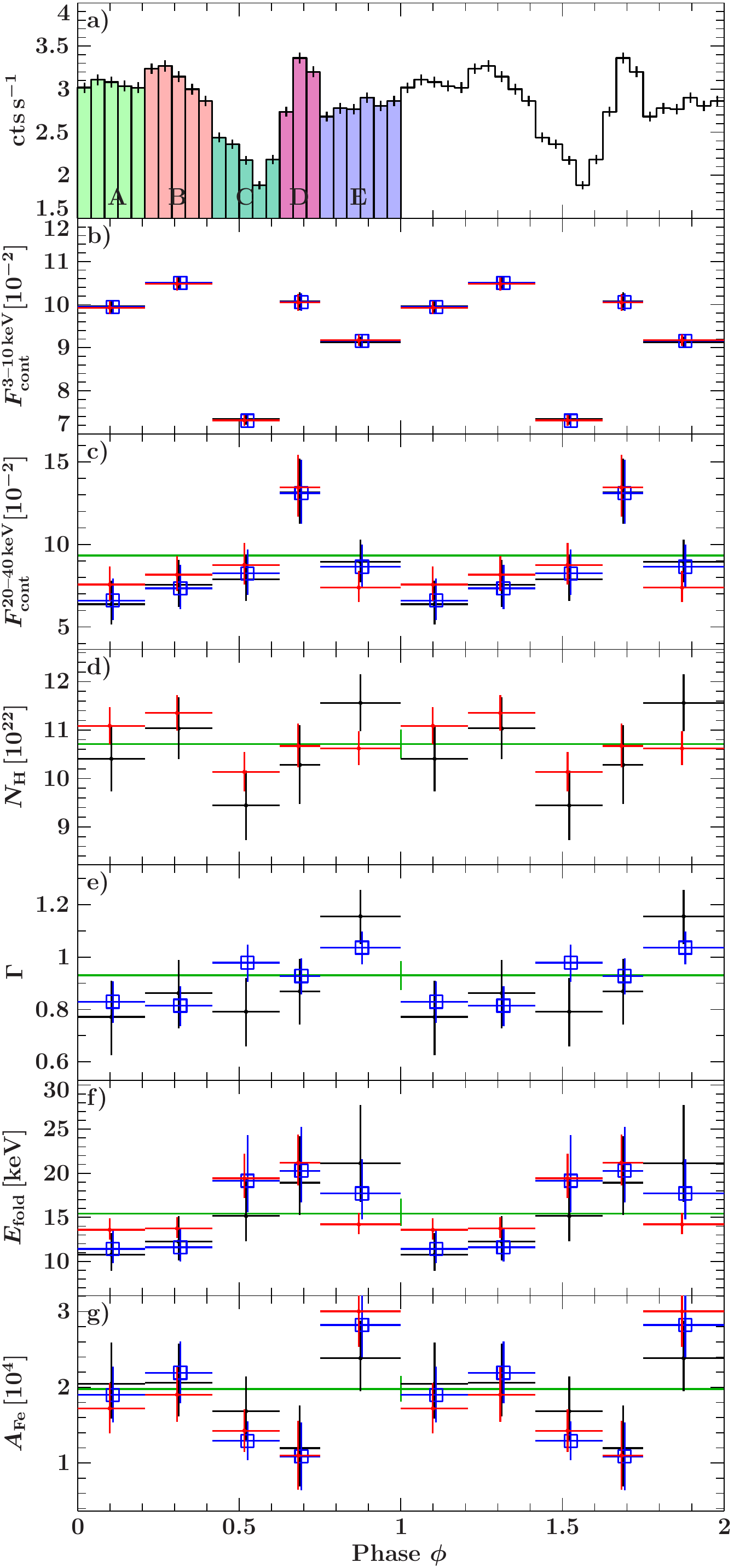}
 \caption{Results of the phase resolved spectroscopy. In black the fit without any fixed parameters is shown, for the blue data points the \nh was frozen, for the red the photon index $\Gamma$. In green the results of the phase averaged fit are shown. \textit{a)} XIS0 pulse-profile in the 0.5--10\,keV range, \textit{b)} flux between 3--10\,keV in keV\,s$^{-1}$\,cm$^{-2}$,  \textit{c)} flux between 20--40\,keV in keV\,s$^{-1}$\,cm$^{-2}$, \textit{d)} absorption column density, \textit{e)} photon index, \textit{f)} folding energy, \textit{g)} flux of the iron line in ph\,s$^{-1}$\,cm$^{-2}$.}
 \label{fig:phasresfit}
\end{figure}

To study the variance of the spectrum with the viewing angle onto the neutron star, i.e., the pulse phase, we divided the data in 5 phase bins, as indicated in Fig.~\ref{fig:ppergres}. The bins were chosen to cover the prominent features, e.g., the deep minimum between the pulses and the hard primary pulse (bins C and D, respectively) as well as to retain a \snr sufficient to constrain the individual spectral parameters well. As shown by \fuerst, the \texttt{cutoffpl} model delivers the best description of the phase-resolved spectra, and in order to be able to compare our results directly with the previous work, we  applied  that model to the phase-resolved data. Neither a blackbody component nor a partial coverer could be significantly detected in these spectra due to the lower \snr, therefore we only use a single absorber, the continuum, and a Gaussian-shaped fluorescent iron line at 6.4\,keV.

Detailed analysis of the phase averaged spectrum shows that there are strong systematic correlations between \nh, $\Gamma$, and $E_\text{fold}$. These correlations lead to large uncertainties in the phase resolved spectra in all parameters. To reduce these uncertainties we fixed each of the three parameters in turn, and compared the results with the energy resolved pulse profiles. All fits lead to very similar \redchi values, so that a distinction based on the quality of the fit is not possible. However, when comparing the fluxes in the 20--40\,keV energy band modeled with a fixed folding energy to the pulse profile in that energy range, a clear discrepancy is evident, i.e., the bright main peak is not correctly reproduced. Additionally, \fuerst found indications that the phase resolved \xte spectra are best described using a variable $E_\text{fold}$. We therefore exclude the  model with a fixed folding energy from the following discussion. 

As can be seen in Fig.~\ref{fig:phasresfit}f, the folding energy is in all models very high during phase bins C and D, the minimum and the primary peak. 
The folding energy seems therefore responsible for the spectral hardening during the primary peak, and not, e.g., strong changes in the photon index $\Gamma$. The results also show that the minimum has a similar spectrum as the primary peak. In the models with variable $\Gamma$, $E_\text{fold}$ stays high during phase bin E, but at the same the time $\Gamma$ is rising, leading to an overall softer spectrum. With $\Gamma$ frozen, $E_\text{fold}$ drops to values similar to those in phase bins A and B. Differences between these two descriptions would only be visible above $\sim$70\,keV, where the source flux has dropped below the detection limit of PIN. 

Variations of \nh are visible in Fig.~\ref{fig:phasresfit}d, but seem uncorrelated with the observed hard primary peak. \nh drops to low values during the minimum in phase bin C, but the overall variation is only marginally significant. It is interesting to note that the flux of \feka is also variable with phase (Fig.~\ref{fig:phasresfit}g), however it is not correlated with the overall soft X-ray flux nor the \nh variations. There rather seems to be a slight shift with the \feka flux reacting on the 3--10\,keV flux with approximately 120\,s delay. Using the model in which all parameters were allowed to vary and shifting the \feka flux by one phase bin, the correlation to the soft continuum increases from 0.04 to 0.70 according to Pearson's correlation coefficient. The energy of the \feka line (not shown) is not varying significantly with pulse phase.

\section{Summary \& Outlook}
\label{sec:summary}
We have presented a timing and spectral analysis of a 20\,ks \suz observation of \fu. We extended the pulse period evolution as presented by \fuerst by two data points, the most recent of which, $P= 604.86$\,s, is consistent between \inte and \suz. To describe the spectrum we applied different phenomenological models and find statistically acceptable fits when including a fluorescent \feka line at 6.4\,keV and either a blackbody component or a partially covering absorber. The continuum parameters in the models with a blackbody component are very similar to the ones found by \fuerst in \xte-data. However, they are highly unconstrained in the \suz-data, while using a partial coverer still results in a good description of the continuum with small uncertainties.

A partial coverer can be used as an approximation to a complex structured stellar wind (``clumpy'' wind), as seen in many HMXBs as well as isolated O and B stars \citep[see, e.g.,][among many others]{oskinova07a, fuerst11b}. Within these winds, dense structures with column densities of the order of $10^{24}$\,cm$^{-2}$ can be reached, as seen in many other sources \citep[e.g., GX~301$-$2 or IGR~J16318$-$4848;][]{fuerst11b,barragan09a}. The absorption column of the partial coverer in \fu is measured to be also of that order, with values between (0.5--1)$\times10^{24}$\,cm$^{-2}$, as seen in Tables~\ref{tab:phasavg_ecu}, \ref{tab:phasavg_hie}, and \ref{tab:phasavg_ctt}.

We also applied the partial covering model to the older \xte data, which have about twice the column density as the \suz data, but did not find an acceptable fit. As the \xte data do not cover the very soft energies, a small contribution of the partial coverer can easily be missed in the data. If the partial coverer is due to the stellar wind, it is also highly variable, as structures will move with the wind with typical outflow terminal velocities  $\geq 1000$\,km\,s$^{-1}$ \citep{prinja90a}. Therefore, the \xte observation might have just been performed during a less obstructed view onto the X-ray source. 
Nonetheless, these data clearly require a blackbody component to describe the data. 

Assuming that the model including the partial coverer represents a good description of the physical conditions in and around the X-ray source, it means that the blackbody disappeared, i.e., that it is highly variable. The blackbody component originates from the thermal mound on the neutron star or the accretion column and its properties can vary with changes in the accretion rate and the physical conditions in the accretion column close to the neutron star. The X-ray fluxes of the \suz and \xte data are, however, similar, i.e., it is likely that the accretion rate and thereby the temperature in the column are also very similar. Changes in the geometry of the accretion column could account for small differences, like the ones seen when comparing the \xte model to the best fit \suz model which includes a blackbody component. In \suz the temperature is measured to be around 2\,keV, compared to $\approx 1.5$\,keV in the \xte data. That the blackbody component becomes completely invisible due to changes in the geometry of the accretion column seems unlikely. 
It is also possible that the blackbody is
a spurious effect in the \xte data. For example, the galactic ridge emission, which can strongly influence \xte/PCA data due to the large field of view of the collimator, can also often be described by a soft thermal component \citep[see, e.g.,][]{yamauchi09a}. The galactic ridge component is suppressed in \suz/XIS due to its imaging capabilities.
To study the variability of the blackbody and the partial coverer, a more extensive monitoring in the soft X-rays is necessary.

To investigate the behavior of the spectrum under different viewing angles, we performed phase resolved spectroscopy in 5 phase bins. As already seen in the energy resolved pulse profiles (Fig.~\ref{fig:ppergres}), the primary peak (labeled D) is distinctly harder than the broad secondary maximum (bins A, B, and E). We found that this is likely due to a reduced folding energy in the secondary maximum. This change can be explained by assuming that accretion happens on both poles of a magnetic dipole field and that each pulse maximum corresponds to one of the poles. Small differences in magnetic field strengths will lead to different geometries and thereby temperatures in the two accretion columns. The hotter the electron plasma the more effectively it can scatter photons to high energies via the inverse Compton effect, leading to an overall harder spectrum. This picture might be over simplified, as relativistic light bending effects and the geometry of the accretion column need to be taken into account, but such an analysis is beyond the scope of this article. There are, however, new models under development to describe the X-ray producing region more realistically \citep[e.g.,][]{becker12a}, which we will apply in a future publication.

We searched for CRSFs in the phase averaged as well as in the the phase resolved spectrum, but did not find any evidence for such a feature in the energy range between 10 and 70\,keV. Assuming typical values of $E_\text{CRSF} = 25$\,keV and $\sigma_\text{CRSF} = 10$\,keV the maximal depth for such a line would be $\tau_\text{CRSF} \leq 0.23$, assuming a line with a Gaussian optical depth. This is roughly a factor of two less than the depths of the lines seen for example in A~0535+269 \citep{caballero07a} or 4U~1907+09 \citep{rivers10a}.
As \fu is clearly a young neutron star, as proven through the strong pulsations and the early type stellar companion, it also possesses a strong magnetic field, which in theory must lead to the formation of CRSFs. As theoretical simulations show, however, it is possible that these features (especially the fundamental line) can be filled up again by photon spawning and are therefore not detectable in the observed X-ray spectrum \citep{schoenherr07a}. Additionally, the geometry of the magnetic field as well as the viewing angle onto the neutron star both can have strong effects regarding line shape and depth. Work is still in progress in understanding and quantizing these effects. The stringent upper limit on the CRSF depth is therefore very important to better understand the physics and magnetic fields of neutron stars.

\acknowledgements
This work was supported by the Bundesministerium f\"ur Wirtschaft und Technologie through DLR grants 50\,OR\,0808, 50\,OR\,0905, and 50\,OR\,1113. FF thanks GSFC for the hospitality. This research has made use of data obtained from the \suz satellite, a collaborative mission between the space agencies of Japan (JAXA) and the USA (NASA). This work is furthermore based on observations with \inte, an ESA project with instruments and science data centre funded by ESA member states (especially the PI countries: Denmark, France, Germany, Italy, Switzerland, Spain), Czech Republic and Poland, and with the participation of Russia and the USA. We have made use of NASA's Astrophysics Data System. We like to thank J.~E.~Davis for the \texttt{slxfig} module which was used to create all plots throughout this paper.

\end{document}